\documentclass[a4paper,12pt]{article}
\usepackage[toc,page]{appendix}
\usepackage{geometry,slashed}
 \usepackage
 {
   amsmath,amssymb,graphicx,tabularx
 }
  \usepackage{soul}
  \usepackage[usenames,dvipsnames]{xcolor}
 
 \definecolor{X575}{rgb}{0.05, 0.7, 0.05}
 
 \usepackage{color}
 \usepackage{hyperref}
 \usepackage{caption,subcaption}
 \usepackage{authblk}
 \usepackage{epstopdf}
 \usepackage{enumerate}
 \usepackage{multirow}
 \usepackage{siunitx}
 \usepackage{float}
  \usepackage{color}
 \usepackage{hyperref}
 \usepackage{caption,subcaption}
 \usepackage{authblk}
 \usepackage{epstopdf}
 \usepackage{enumerate}
 \usepackage{multirow}
 \usepackage{siunitx}
 \usepackage{float}
 \usepackage{cancel}
 \usepackage{multicol}
 \usepackage{cite}
 \usepackage{comment}

  \usepackage{listings}
  \usepackage{calrsfs}

  \usepackage{listings}
  \usepackage{calrsfs}
  \usepackage{adjustbox}
  \DeclareMathAlphabet{\pazocal}{OMS}{zplm}{m}{n}

\title{Probing the spin correlations of $t\bar t $ production at NLO QCD+EW}

\author[]{Rikkert Frederix\thanks{rikkert.frederix@thep.lu.se}}
\author[]{Ioannis Tsinikos\thanks{ioannis.tsinikos@thep.lu.se}}
\author[]{Timea Vitos\thanks{timea.vitos@thep.lu.se}}

\affil[]{\small Theoretical Particle Physics, Department of Astronomy and Theoretical Physics, Lund University, S\"olvegatan 14A, SE-223 62 Lund, Sweden}

\begin{document}
\date{}
\maketitle
 
\vspace*{-8cm}
{
     {LU-TP 21-13}
}
\vspace*{8cm}

\begin{abstract}
\noindent
In this work we investigate the NLO QCD+EW corrections to the top quark pair production and their effects on the spin correlation coefficients and asymmetries at fixed-order top quark pair production and LO decay in the dilepton channel, within the narrow-width approximation. The spin correlations are implicitly measured through the lepton kinematics. Moreover we study the EW effects to the leptonic differential distributions. We find that the EW corrections to the $t \bar t$ production are within the NLO QCD theoretical uncertainties for the spin correlation coefficients and the leptonic asymmetries. On the other hand, for the differential distributions we find that the EW corrections exceed the NLO QCD scale uncertainty band in the high rapidity regimes and are of the order of the NLO QCD scale uncertainty in the case of invariant mass and transverse momentum distributions.
\end{abstract}

\tableofcontents

\section{Introduction}

As the heaviest particle in the well-established Standard Model of particle physics, the top quark is of interest for investigation for multiple purposes. Firstly, its role in the fine-tuning problem is exquisite. Secondly, it is a close portal to any beyond Standard Model theories. To make precise predictions for observables related to top quarks is of utmost interest for understanding the present idea of particle physics and to enlarge our views of it by extensions to the existing theories. \\

\noindent
After its discovery in 1995, the top quark has been detected in numerous high-energy processes, both in single-top, top quark pair, and production in association with other particles. In pair production, the spin correlation effects serve as a probe for investigation of the underlying physics. Due to its high mass and hence  short lifetime, the top quark decays (with almost 100~\% branching ratio to $Wb$) before the spin decorrelation sets in. Thus, the decay products of top quarks carry the spin information and hence provide an optimal source for understanding the nature of top quarks by indirect spin correlation measurements. The prominent decay channel is the all-hadronic channel, in which both the weak bosons decay to quarks. However, the dilepton decay channel (with branching ratio $\sim 10.5\%$) provides a cleaner signature at hadron-hadron collisions and thus an important choice for investigation on the spin correlation effects. \\

\noindent
The top quark spin correlations and the differential distributions of the decay products have been previously studied at various levels of precision. The effects of the top quark spin correlations are revealed to the dilepton decay channel. They are studied at NLO QCD accuracy \cite{Bernreuther:2003ga, Bernreuther:2001rq, Bernreuther:2001bx}, at NLOW (NLO QCD + Weak) accuracy \cite{Bernreuther:2013aga} and at NLOW through the spin correlation coefficients \cite{Bernreuther:2015yna}. The spin correlation coefficients have been compared to experimental data at both ATLAS \cite{Aaboud:2016bit} and CMS \cite{CMS:2018jcg} collaborations. In the latter case the lepton asymmetries are also presented and furthermore a comparison of the spin-correlated and uncorrelated NLO QCD predictions is done to point out the importance of including the spin correlations in calculations of such observables. A special attention was brought to one observable,  $\Delta\Phi_{\ell \ell}$, the angle in the transverse plane of the charged leptons, which was measured by CMS \cite{CMS:2018jcg} and ATLAS \cite{Aad:2019hzw} and compared to NLO QCD predictions. This was followed by the NNLO QCD study \cite{Behring:2019iiv}, where the specific observable was found to deviate from the Standard Model predictions in the inclusive (unfolded measurements)  region, while fitting nicely to the theory predictions in the fiducial region. The decay product differential distributions for the top quark pair semi-leptonic and leptonic decay are studied at NLO QCD with the top quarks on-shell \cite{Melnikov:2009dn}, and including off-shell effects and non-resonant contributions \cite{Denner:2017kzu,Jezo:2016ujg}. The dilepton channel is studied including part of the NLO EW corrections \cite{Denner:2016jyo}. The leptonic differential distributions are also compared to the experimental data in ATLAS \cite{Aad:2019hzw} and CMS \cite{Sirunyan:2018ucr}. Recently in Ref. \cite{Czakon:2020qbd}, all the aforementioned observables are calculated at NNLO QCD with on-shell top quark decays and the results are compared with experimental data. Despite the high level of accuracy, there are still slight tensions between theory and experiment in the normalised differential distributions, like the $\Delta\Phi_{\ell \ell}$ at the inclusive level, the invariant mass and transverse momentum of the lepton pair, as well as the transverse momentum of the leptons in the fiducial region. The different level of agreement of the $\Delta\Phi_{\ell \ell}$ distribution with the data in the fiducial and the inclusive regime shows that most probably the reason is the use of NLO QCD predictions for the unfolding to the full phase space. This is discussed in detail in Ref. \cite{Behring:2019iiv}. Furthermore in Ref. \cite{Czakon:2020qbd} it is shown that the tension of this observable with the data in the inclusive phase space persists even after using the expanded definition at NNLO in QCD. Regarding  other leptonic distributions ($e.g.$ $m(\ell \ell), p_T(\ell \ell)$), slight tensions are shown in Ref. \cite{Czakon:2020qbd} in the ratios of the data w.r.t. the NNLO QCD predictions in the fiducial phase space. \\

\noindent
As no clear explanation to these slight discrepancies between theory and data has yet been found, a natural next step in the investigation within the Standard Model is the inclusion of the full EW effects at NLO to a specific set of observables. In this work we calculate at fixed-order, for the first time, the complete NLO $t \bar t$ production followed by the LO top quark pair decays through the dilepton channel in the narrow-width approximation (NWA). We focus our study on the spin correlation coefficients, the leptonic asymmetries and normalised differential distributions. We further explain the reweighting technique, which is used in the decay chain.  \\

\noindent
The structure of this paper is as follows. In Sec. \ref{sec:theory} we introduce the theoretical framework in which we calculate the observables perturbatively, then we introduce the reweighting procedure for fixed-order phase-space points, followed by the spin correlation coefficient and asymmetry definitions. In Sec. \ref{sec:numerics} we explain our calculational setup and input parameters. In Sec. \ref{sec:results} we present our results for the coefficients and asymmetries, and a set of various distributions in the  leptonic kinematics, in all cases comparing NLO QCD and complete-NLO results. Finally, we discuss our results and conclude in Sec. \ref{sec:conclusion}.

\begin{comment}
Besides the set of spin correlation coefficients which were introduced in Ref.~\cite{Bernreuther:2001rq,Czakon:2020qbd}, another interesting set of observables which probe the spin properties of the particles are the induced asymmetries. These factors measure an asymmetry in the direction of flight of decay products, being a remnant of spin properties of the contributing particles in combination with other effects. The asymmetry factors depend largely on the type of hadron-hadron collider: Tevatron results and LHC results \cite{Aad:2015jfa,Sirunyan:2019eyu} differ largely, however both show clear evidence of  some level of observed asymmetry. Calculations beyond NLO QCD for the asymmetries in top quark pair production have been performed previously in Ref. \cite{Ahrens:2011uf}.\\
\end{comment}

\section{Theoretical setup}\label{sec:theory}

The dilepton decay channel of top quark pair production has contributions from non-resonant, single-resonant and double-resonant diagrams. In the NWA, in which the top quarks are produced on-shell, only the double-resonant diagrams are taken into consideration. A full calculation of the dilepton final state $\ell^+ \ell^- \nu\overline{\nu} b \overline{b}$ would cover all these contributions to this process. However, the non-double-resonant contributions are expected to be very small for the observables studied in this work, due to the small width-to-mass ratio $\Gamma_t/m_t$ of the produced top quarks. Within the decay chain, resonant particles are produced on-shell and in a next step are further decayed, in this way compactly implementing the NWA. \\

\noindent
Utilising MadGraph5\_aMC@NLO \cite{Frederix:2018nkq} and its internal tool for decaying resonant particles, MadSpin \cite{Artoisenet:2012st}, one efficiently achieves a fully-decayed set of events. The current version of MadSpin calculates all decays and the corresponding spin correlations at tree-level. Assigning these decays to a set of production phase-space (PS) points from leading-order generation is trivial, as there is no ambiguity in the way the PS points are reweighted. The algorithm for obtaining decayed PS points from a fixed next-to-leading order sample is however currently not yet implemented in the public version of MadSpin. For the current project where the NLO effects, including both the QCD and EW corrections, are included in the production, a method for reweighting fixed-order PS points is developed and utilised. \\

\noindent
It should be noted, however, that MadSpin does not handle the virtual corrections in the production PS points and hence these are not included in the spin correlations. To properly describe this, let us first introduce the notation for the perturbative expansions of the process  $pp \rightarrow t \overline{t}$. We expand a general observable $\Sigma$ at fixed-order in the electroweak $\alpha$ and strong $\alpha_S$ coupling constants with the following notation,
\begin{eqnarray}\label{Eq:orders}
\begin{split}
\Sigma_{\text{LO}}(\alpha,\alpha_S) &=\underbrace{ \alpha_S^2 \Sigma_{2,0}}_{\text{LO$_1$}} + \underbrace{\alpha_S \alpha \Sigma_{1,1}}_{\text{LO$_2$}} + \underbrace{\alpha^2 \Sigma_{0,2}}_{\text{LO$_3$}} \\
\Sigma_{\text{NLO}}(\alpha,\alpha_S) &= \underbrace{\alpha_S^3 \Sigma_{3,0}}_{\text{NLO$_1$}}+ \underbrace{\alpha_S^2\alpha \Sigma_{2,1}}_{\text{NLO$_2$}}+  \underbrace{\alpha_S\alpha^2 \Sigma_{1,2}}_{\text{NLO$_3$}}+\underbrace{\alpha^3 \Sigma_{0,3}}_{\text{NLO$_4$}}
\end{split}
\end{eqnarray}
where the NLO pieces  include both the virtual and real corrections. We refer to the leading-order term of LO$_1$ as LO QCD, the leading-order NLO contribution of LO$_1$+NLO$_1$ as NLO QCD and the leading NLO electroweak contribution of LO$_1$+NLO$_2$ as NLO EW. In the present work, the NLO QCD+EW refers to the complete NLO, that is, the full LO$_1$+LO$_2$+LO$_3$ +NLO$_1$+NLO$_2$+NLO$_3$+NLO$_4$. Regarding the full process under study we attach the LO leptonic top quark pair decay ($O(\alpha^4$)) to the perturbative orders in Eq.~\ref{Eq:orders}. \\

\noindent
We extend MadSpin to include decays of fixed-order PS points. MadSpin reads the production LHE file with the fixed-order PS points and re-computes the tree-level matrix element for each PS point including the decay, approximating the virtual diagram contributions with the tree-level ones. Top quark pair production has been also studied at NLO QCD in production and decay within the MCFM framework \cite{Campbell:2012uf}. We will use this tool to examine the effects of the spin correlations from the virtual NLO QCD part of the $t \bar t$ production.

\subsection{Reweighting in fixed-order PS point generation}

Reweighting in the decay chain approximation from uncorrelated to spin-correlated decays was developed for event generation with parton showering in Ref.~\cite{Frixione:2007zp}. The method is based on relating the differential weight of the fully decayed event to the production event weight. In this manner, one may show, as is done in Ref.~\cite{Frixione:2007zp}, that the fully decayed event weight is bounded from above by 
\begin{eqnarray}
\frac{\text{d}\sigma}{\text{d}(\Omega_{\text{full}})}< B_{\rm max} \frac{\text{d}\sigma}{\text{d}(\Omega_{\text{prod}})},
\label{eq:rmax}
\end{eqnarray}
where $B_{\rm max}$ is a process-dependent calculable (within perturbation theory) quantity. In practice, this upper bound, denoted by $r_{\rm max}$, is obtained by probing an adequate sized subsample of production PS points with decays and extracting the maximum value. \\

\noindent
For fixed-order generation, the PS points are stored in an LHE file format, with \texttt{eventgroup} labeling the groupings of PS points which cancel in the IR regions \cite{Butterworth:2010ym}. In the top quark pair production, the PS points consist of a $2 \rightarrow 2$ configuration (we refer to these as Born PS points with subscript $B$, but include also virtual and soft and/or collinear counter terms), with two top quarks in the final state, and $2 \rightarrow 3$ configurations (we refer to these as real PS points, with subscript $R$), with top quarks plus an additional real emission particle  in the final state. Let $N$ be the number of eventgroups in the LHE production file, let $d_{B/R}$ denote the weight of the fully decayed PS point for the Born or real PS points, and $p_{B/R}$ denote similarly the weights at production stage. Schematically, the cross section for the fully decayed process $\sigma_{\rm dec}$ can then be rewritten as
\begin{eqnarray}\label{eq:reweight}
\sigma_{\rm dec} = \sum_{i=1}^N (d_B^i+d_R^i)=r_{\rm max}\sum_{i=1}^N \frac{r_B^i}{r_{\rm max}}\left(p_B^i + \frac{r_R^i}{r_B^i}p_R^i  \right)
\end{eqnarray}
where we define the ratios $r_{B/R}^i$ as the ratios between the fully decayed and production weights, 
\begin{eqnarray}
r_{B/R}^i = \frac{d_{B/R}^i}{p_{B/R}^i},
\end{eqnarray}
and $r_{\rm max}$ is the upper bound as we have defined it right after Eq.~\ref{eq:rmax}, obtained from the Born configurations. Now the procedure is to note that the PS point weight may be decomposed into, following notation from Ref.~\cite{Mattelaer:2016gcx}
\begin{eqnarray}
p^i = f_ {1/h_1}(x_1^i,\mu_F)f_ {2/h_2}(x_2^i,\mu_F) |M_{\rm prod}^i|^2 \text{d}\Omega^{\rm prod}_i
\end{eqnarray}
where $f_{i/h_j}(x,\mu_F)$ denotes the parton distribution function for parton $i$ in hadron $j$ at the longitudinal momentum fraction $x_i$ at factorization scale $\mu_F$. $|M^i|^2$ denotes the (color- and spin-summed) matrix-element squared for the given PS point and finally $\text{d}\Omega^{\rm prod}_i$ is the corresponding phase-space factor (including flux factors) in the production final state. The fully decayed PS point weight is similarly written as 
\begin{eqnarray}
d^i = f_ {1/h_1}(x_1^i,\mu_F)f_ {2/h_2}(x_2^i,\mu_F) |M_{\rm full}^i|^2 \text{d}\Omega^{\rm full}_i.
\end{eqnarray}
Attachment of an unweighted decay PS point to a production PS point retains the same parton distribution functions and hence these cancel in the ratio 
\begin{eqnarray}\label{eq:ratio}
r^i = \frac{|M_{\rm full}^i|^2 \text{d}\Omega^{\rm full}_i}{ |M_{\rm prod}^i|^2 \text{d}\Omega^{\rm prod}_i},
\end{eqnarray}
where we omit the subscript $B/R$. The full phase space factorises into the production and decay phase space,
\begin{eqnarray}
\text{d}\Omega_{\rm full} \propto \text{d}\Omega_{\rm prod} \text{d}\Omega_{\rm dec},
\end{eqnarray}
and for an unweighted sample of decay PS points, the decay phase space gets mapped to 
\begin{eqnarray}
\text{d}\Omega_{\rm dec} \rightarrow  \frac{\text{d}\Omega_{\rm dec}}{|M_{\rm dec}|^2},
\end{eqnarray}
where we have omitted any constants including the flux factor, total branching ratio, and number of PS points. Finally, this allows us to write the reweighting ratio in Eq.~\ref{eq:ratio} as 
\begin{eqnarray}\label{eq:me}
r^i \propto \frac{|M_{\rm full}^i|^2 }{|M_{\rm prod}^i|^2|M_{\rm dec}|^2}.
\end{eqnarray}
For its purpose of use, in Eq.~\ref{eq:reweight}, in the ratio $\frac{r_R^i}{r_B^i}$ the constants drop out, leaving the relevant piece to be fully determined by the amplitudes squared of the production, fully decayed and decay PS point only as given by Eq.~\ref{eq:me}.  \\

\noindent
For exact IR pole cancellations, in the soft and collinear regions, this reweighting step must yield exactly identity factors, $\frac{r_R^i}{r_B^i} \rightarrow 1$. Due to numerical imprecision, this ratio of ratios will yield cases in which they are different from unity with a significant amount. Hence, the final cross section and distributions are prone to large statistical errors when these cancellations are not exact. To remedy this numerical inaccuracy, we introduce a smooth mapping function, which maps the ratios to the form 

\begin{eqnarray}
\frac{r_R^i}{r_B^i} = \begin{cases} 
1  &\text{for } \; p_T < x_{\rm min} \\
1 + D(p_T) \left(\frac{r_R^i}{r_B^i}-1\right) &\text{for }\;  x_{\rm min} \le p_T \le x_{\rm max} \\
\frac{r_R^i}{r_B^i}  &\text{for }\; p_T > x_{\rm max}
\end{cases}
,
\end{eqnarray}

\noindent
where $p_T$ is the transverse momentum of the extra emission\footnote{Formally this function introduces a small dependence on the FKS mapping~\cite{Frederix:2009yq,Frixione:1995ms} for the values of $p_T < x_{\rm max}$. However in practice this effect is expected to be negligible.}. The smooth damping function $D(x)$ is chosen arbitrarily. In our calculations we use the smooth step function to third order,
\begin{eqnarray}
D(x) = 3 \frac{(x-x_{\rm min})^2}{(x_{\rm max}-x_{\rm min})^2}-2\frac{(x-x_{\rm min})^3}{(x_{\rm max}-x_{\rm min})^3}
\end{eqnarray}
\noindent
with $x_{\rm min},x_{\rm max}$ chosen empirically. Thus, the values used for the calculation, $x_{\rm min} =15.0$ GeV  and $x_{\rm max}=30.0 $ GeV, were obtained by comparing a set of values and picking the two limits such that the numerical fluctuations vanish, at the same time as the interval of this soft damping $x_{\rm max}-x_{\rm min}$ is chosen to be as small as possible. It should be noted that for massless final states, the $p_T$ of the extra emission cannot act as a proxy for the soft and collinear regions simultaneously as it does for the present case of massive top quarks. In general, one must introduce a check for all the collinear regions carefully, and in that way obtain the smooth mapping function which smoothens the boundary for the soft/collinear regions.

\subsection{Spin correlation coefficients}
\label{sec:spin_corr_coeff}

A thorough examination of spin correlations for top quark pair production is done through the spin-density formalism \cite{Bernreuther:1993hq,Bernreuther:2015yna}. This introduces a set of coefficients which parameterize the cross section along different axes in some reference frame in which the spins of the top quarks are expressed. In order to present this, we must first introduce the frame of reference in which we define the leptonic angles. \\

\begin{figure}[htb!]
\center
\includegraphics[scale=0.18]{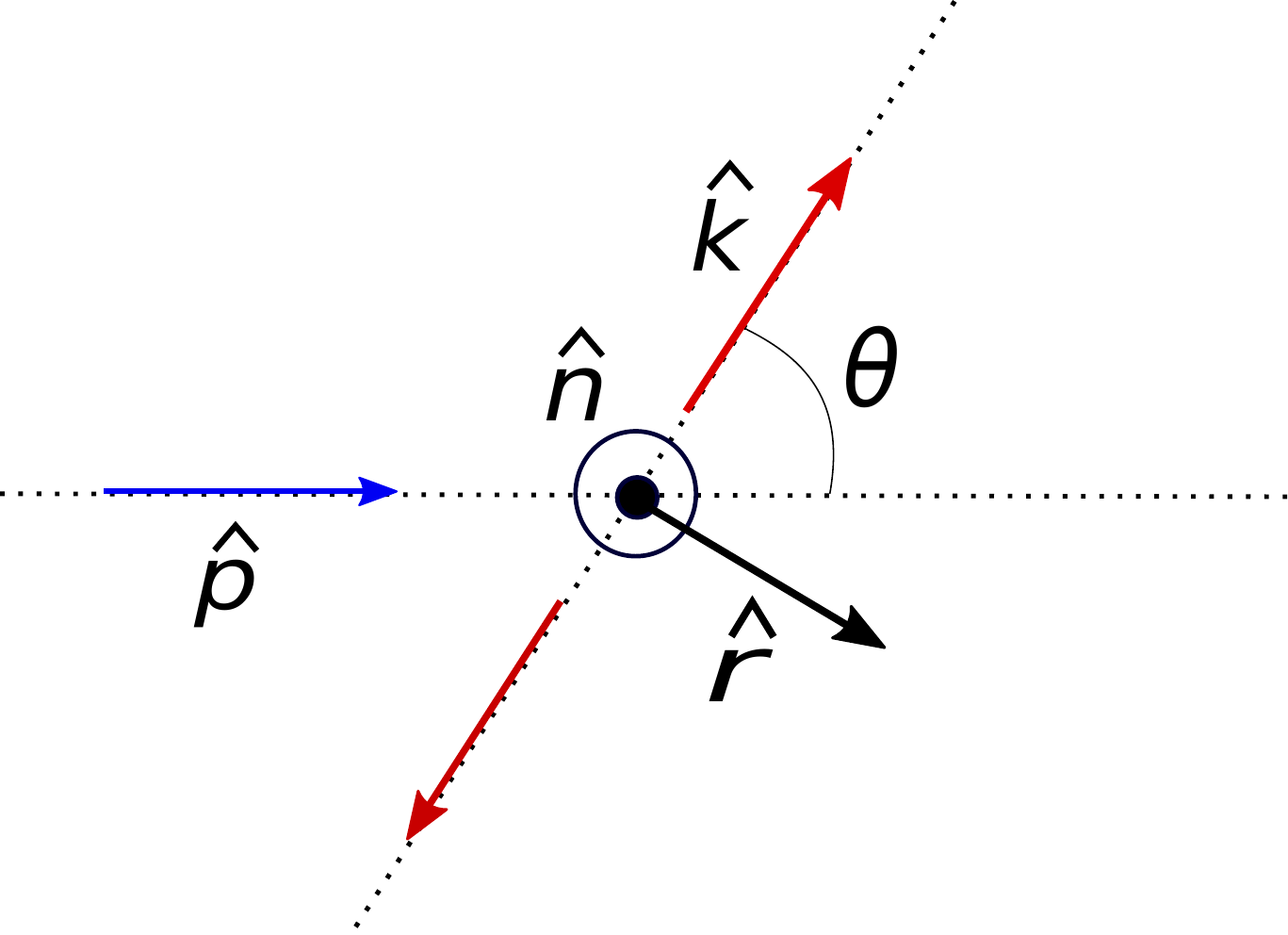}
\caption{The orthonormal basis for the spin projections. The blue arrow indicates the direction of one of the proton beams in the laboratory frame. The red arrows indicate top and anti-top directions in the $t\overline{t}$ center of momentum frame. The angle $\theta$ defines the angle between the beam and the outgoing top quark.}
\label{fig:rkn}
\end{figure}

\noindent
One introduces a suitable orthonormal basis for the spin directions. We follow the definitions and notations used in Ref.~\cite{Bernreuther:2015yna}. Let $\hat{k}$ be the direction of the top quark in the $t\overline{t}$ center of momentum frame, rotation-free boosted from the laboratory frame, and $\hat{p}$ be the direction of motion of one of the proton beams in the laboratory frame. We then define a normal $\hat{n}$ to the plane spanned by $\hat{k}$ and $\hat{p}$ and finally define a unit vector $\hat{r}$ which completes to a right-handed orthonormal basis $\{\hat{k},\hat{n},\hat{r}\}$, see Fig. \ref{fig:rkn}, given by
\begin{eqnarray}
\begin{split}
\hat{n} &= \text{sign}(\cos \theta) \frac{1}{\sin{\theta}}\hat{p}\times \hat{k},\\ \hat{r}&= \text{sign}(\cos\theta)\frac{1}{\sin \theta} (\hat{p}-\hat{k}\cos \theta),
\end{split}
\end{eqnarray}
where the $\text{sign}(\cos\theta)$ is introduced in order to maintain non-vanishing coefficients due to CP transformations \cite{Bernreuther:2015yna,Czakon:2017lgo}. Then finally we are ready to define the leptonic angles in this frame,
\begin{eqnarray}
\begin{split}
\cos\theta_+^k &= \hat{p}_{\ell^+} \cdot \hat{k} \quad \quad \quad \quad \cos\theta_-^k = \hat{p}_{\ell^-} \cdot \hat{k} \\
\cos\theta_+^n &= \hat{p}_{\ell^+} \cdot \hat{n} \quad \quad  \quad \quad \cos\theta_-^n = \hat{p}_{\ell^-} \cdot \hat{n}  \\
\cos\theta_+^r &= \hat{p}_{\ell^+} \cdot \hat{r} \quad \quad  \quad \quad \cos\theta_-^r = \hat{p}_{\ell^-}\cdot \hat{r}, 
\end{split}
\end{eqnarray}
with the charged lepton directions of motion $\hat{p}_{l^\pm}$ and the angles are defined to be $\theta_+$ for the $l^+$ and $\theta_-$ for the $l^{-}$. In terms of these definitions, the normalised differential distribution is expanded according to

\begin{eqnarray}
\begin{split}
\frac{1}{\sigma} \frac{d\sigma}{d \cos \theta_+^i d \cos \theta_-^j} = \frac{1}{4} \left( 1+B_+ ^i  \cos \theta_+^i + B_-^j \cos \theta_- ^j 
 + C_{ij} \cos \theta_+^i \cos \theta_-^j \right) 
\end{split}
\end{eqnarray} 
for $i,j = \{\hat{k},\hat{n},\hat{r}\}$. This introduces a set of six $B$ coefficients  and nine $C$ coefficients. By expansion and integration of this expression, one may extract the coefficients from a fixed-order PS point generation by the identities
\begin{eqnarray}\label{eq:coeffdef}
C_{ij} = -9 \frac{\langle\cos \theta_+^i \cos \theta_- ^j\rangle}{\sigma}, \quad \quad B_\pm^i =  3 \frac{\langle\cos \theta_\pm^i\rangle}{\sigma}
\end{eqnarray}
in which the factors $-9$ and $3$ are introduced to maintain correct normalisation \cite{Bernreuther:2015yna}.\\

\noindent
For all coefficients $\mathcal O = B,C$, the perturbative expansion in the QCD coupling may be obtained via an unexpanded or an expanded expression. For both of these groups of coefficients, we may write the functional form as (absorbing in the numerator the numerical factors of 3 and -9)

\begin{eqnarray}\label{eq:quotient}
\mathcal O = \frac{\sigma_{_{\mathcal{O}}}}{\sigma} = \frac{\sigma_{\mathcal{O}, \rm NLO}^{\rm QCD+EW}}{\sigma_{\rm NLO}^{\rm QCD+EW}} \; .
\label{eq:coeff}
\end{eqnarray}

\noindent
This definition of the observables is a perturbative series in the strong coupling $\alpha_S$ and the electroweak coupling $\alpha$. The unexpanded definition results in the evaluation of the numerator and denominator separately at the order of interest followed by the division. An expanded version corresponds to the explicit expansion in $\alpha_S$ of the quotient function. In the following we denote with $N$ and $D$ the numerator and denominator functions respectively, with the subscripts denoting the order of accuracy.\\

\noindent
The NLO QCD accurate quotients, unexpanded, is then 

\begin{equation}
\mathcal O_{\rm NLO}^{\rm QCD} = \frac{N_{\rm NLO}^{\rm QCD}}{D_{\rm NLO}^{\rm QCD}} .
\label{eq:unexp_NLO}
\end{equation}

\noindent
We obtain the expanded version by expanding in $\alpha_S$ and neglecting terms of order $O(\alpha_s^2)$ and higher,
\begin{equation}
\tilde {\mathcal O}_{\rm NLO}^{\rm QCD} = K_{\rm NLO}^{\rm QCD} {\mathcal O}_{\rm NLO}^{\rm QCD}+(1-K_{\rm NLO}^{\rm QCD})\frac{N_{\rm LO}^{\rm QCD}}{D_{\rm LO}^{\rm QCD}} \;,
\label{eq:exp_NLO}
\end{equation}

\noindent
where $K_{\rm NLO}^{\rm QCD}$ is the inclusive NLO QCD $K$-factor. That is, Eq.~\ref{eq:exp_NLO} is a result of an exact $\alpha_S$ expansion of Eq.~\ref{eq:unexp_NLO} up to NLO accuracy. We use this equation as a starting point to define the expanded $\tilde {\mathcal O}$ with EW corrections, assuming that their effects to both the denominator $D$ and the inclusive $K$-factor are negligible. This assumption is valid since these quantities are constructed from the total cross sections. The same strategy is followed in Refs.~\cite{Czakon:2017lgo,Czakon:2014xsa} for the LHC and Tevatron $t \bar t$ asymmetries. This results in the following expression for the complete-NLO expanded observables

\begin{equation}
\tilde {\mathcal O}_{\rm NLO}^{\rm QCD+EW} = K_{\rm NLO}^{\rm QCD} {\mathcal O}_{\rm NLO}^{\rm QCD+EW}+(1-K_{\rm NLO}^{\rm QCD}) \frac{N_{\rm LO}^{\rm QCD}}{D_{\rm LO}^{\rm QCD}} \;.
\label{eq:exp_NLOEW}
\end{equation}

\noindent
Eqs.~\ref{eq:exp_NLO} and \ref{eq:exp_NLOEW} are also used to extract the scale and PDF uncertainties for the expanded observables, after calculating each term for all the separate renormalisation/factorisation scale and PDF variation-member values.

\subsection{Asymmetries}
\label{sec:asymmetry}

The top quark charge asymmetry has been previously calculated for LHC (central-peripheral) and Tevatron (forward-backward) at NNLO QCD + NLO EW in Refs.~\cite{Czakon:2017lgo,Czakon:2014xsa}. In this work, we perform an evaluation of this asymmetry as a cross check of the implementation of the reweighting at fixed-order,  and present novel results for the asymmetries related to the decay products. As customary, we define the rapidity (pseudorapidity) differences for the massive top pair (massless charged lepton pair) as
\begin{align}
\Delta y_{tt} &= |y_t| - |y_{\overline{t}}|,\\
\Delta \eta_{\ell \ell} &= |\eta_{\ell^+}| - |\eta_{\ell^-}|.
\end{align}
\noindent
We define in addition the angular differences for the lepton pair
\begin{align}
\Delta\Phi_{\ell \ell} &= \Phi(\ell^+) - \Phi(\ell^-), \\
\Delta \theta_{\ell \ell} &= \theta(\ell^+) - \theta(\ell^-)
\end{align} 
 with $\Delta\Phi_{\ell \ell}$ being the difference of the lepton azimuthal angles in the transverse plane, and $\Delta \theta_{\ell \ell}$ the angular difference between the charged leptons in the laboratory frame. Below we define the four asymmetries which are examined in the present work.
\begin{itemize}
\item Top central-peripheral asymmetry: 
\begin{eqnarray}
A_{C}^{tt} = \frac{\sigma(\Delta y_{tt}>0)-\sigma(\Delta y_{tt}<0)}{\sigma(\Delta y_{tt}>0)+\sigma(\Delta y_{tt}<0)}
\end{eqnarray}
\item Lepton central-peripheral asymmetry: 
\begin{eqnarray}
A_{C}^{\ell \ell} = \frac{\sigma(\Delta\eta_{\ell \ell}>0)-\sigma(\Delta\eta_{\ell \ell}<0)}{\sigma(\Delta\eta_{\ell \ell}>0)+\sigma(\Delta\eta_{\ell \ell}<0)}
\end{eqnarray}
\item Lepton angular asymmetries: 
\begin{eqnarray}
A_{\Delta\Phi} = \frac{\sigma(|\Delta\Phi_{\ell \ell}|>\frac{\pi}{2})-\sigma(|\Delta\Phi_{\ell \ell}|<\frac{\pi}{2})}{\sigma(|\Delta\Phi_{\ell \ell}|>\frac{\pi}{2})+\sigma(|\Delta\Phi_{\ell \ell}|<\frac{\pi}{2})},
\end{eqnarray}
\begin{eqnarray}
A_{\Delta\theta} = \frac{\sigma(\cos\Delta\theta_{\ell \ell} >0)-\sigma(\cos\Delta\theta_{\ell \ell} <0)}{\sigma(\cos\Delta\theta_{\ell \ell} >0)+\sigma(\cos\Delta\theta_{\ell \ell} <0)}
\end{eqnarray}
\end{itemize}

\noindent
Despite the fact that the numerators of the asymmetries are differently defined with respect to the numerators of the spin correlation coefficients of Sec. \ref{sec:spin_corr_coeff}, as observables they are fully described by Eq.~\ref{eq:coeff}. Therefore the expanded and unexpanded definitions of the asymmetries are described by Eqs.~\ref{eq:unexp_NLO}--\ref{eq:exp_NLOEW}. A simplification occurs in the 
$A_{C}^{tt}$ case, since it is zero at LO QCD, therefore the second term in the expanded definitions drops out. The top quark pair central-peripheral asymmetry was previously calculated in Ref.~\cite{Czakon:2017lgo} at NNLO QCD and NLO EW. We compare our results for this asymmetry obtained from the fixed-order MadSpin module. The asymmetries based on the leptonic kinematics are novel calculations at NLO QCD+EW and hence we make a quantitative comparison to our values obtained at NLO QCD.

\section{Numerical setup}
\label{sec:numerics}

For the numerical calculations we use the 5-flavour scheme NNPDF3.1\_NLO\_luxqed PDF set \cite{Bertone:2017bme} from the LHAPDF library \cite{Buckley:2014ana}, implementing more accurately the photon induced processes for the electroweak corrections. For the fixed-order matrix-element generation and PS point generation, MadGraph5\_aMC@NLO is used. The top quarks, which are produced on-shell, are decayed with MadSpin, which utilises the  tree-level decay chain, based on the NWA. \\

\noindent
The calculation is performed for a 13 TeV center of momentum energy proton-proton collider. The process $p p \rightarrow t \overline{t} \rightarrow W^+ b W^- \overline{b} \rightarrow e^+ \mu^- \nu_e \overline{\nu}_{\mu} b \overline{b}$ is considered, without any jet requirements. In MadSpin, the fixed-order on-shell mode \cite{Mattelaer:2016ynf} is used for this process with the full syntax

\begin{verbatim}
set fixed_order true
set spinmode onshell
decay t > w+ b, w+ > e+ ve
decay t~ > w- b~, w- > mu- vm~
launch
\end{verbatim}
\begin{comment}
For the study of the top pair asymmetry factor, the gluon luminosity in the gluon-gluon decay channel is set (up to numerical precision) to zero, reducing statistics in the symmetric gluon-gluon fusion process of top quark pair production. This reduction of the roughly 85\% production channel allows for a more accurate determination in the numerator of the top quark asymmetry factor (the gluon-gluon contributions canceling for the forward- and backward constellation), while keeping their contribution in the denominator for the total cross section.
\end{comment}

\noindent
For the renormalisation and factorisation scale central values, we use the ones introduced in Ref.~\cite{Czakon:2016dgf} and also used in Ref.~\cite{Behring:2019iiv}, 
\begin{eqnarray}
\mu_R^0=\mu_F^0=\frac{H_T}{4}=\frac{\sum_i m_{T,i}}{4}, \{i=t,\bar t\} \;.
\end{eqnarray}
For the scale variation, the conventional 9-point envelope of $\{ \frac{1}{2}\mu_{R,F}^0,2\mu_{R,F}^0 \}$ is used. The input masses are 
\begin{align}
\begin{split}
&m_{Z}=91.1876~\textrm{GeV}\;,\; m_{W}=80.385~\textrm{GeV}\;,\; \\ &m_H=125~\textrm{GeV}\;,\; m_{\textrm{t}}=172.5~\textrm{GeV} \;.
\end{split}
\end{align}
The widths used are calculated internally by MadSpin, with the values entering the calculations being\footnote{These widths are computed at LO accuracy, consistent with the fact that MadSpin includes the decays at leading order.}
\begin{eqnarray}
\begin{split}
&\Gamma_{Z}=2.44412~\textrm{GeV}\;,\; \Gamma_{W}=2.04542~\textrm{GeV}, \\ &\Gamma_H=4.07468~\textrm{MeV}\;,\; \Gamma_{\textrm{t}}=1.48060~\textrm{GeV}.
\end{split}
\end{eqnarray}
We utilise the $G_{\mu}$ renormalisation scheme in which we use the EW parameters:
\begin{align}
G_\mu=1.166379\times 10^{-5}~\textrm{GeV}^{-2} \Longleftrightarrow \alpha_{\rm EW} = 1/132.2332 \, .
\end{align}

\begin{table*}[htb!]
\begin{center}
\renewcommand{\arraystretch}{1.5}
\scriptsize
\begin{adjustbox}{width=1\textwidth}
\begin{tabular}{c c @{\hskip 20pt} r l @{\hskip 20pt} r l }
\hline
 & & \multicolumn{2}{c}{Unexpanded}  &  \multicolumn{2}{c}{Expanded}  \\
 &  LO QCD [\%] & NLO QCD [\%]  & NLO QCD+EW [\%]  & NLO QCD [\%]  & NLO QCD+EW [\%]  \\
 \hline
$B_k^+$ & $0.01(2)_{-5.0 \%}^{+3.7 \%}~_{-9.8 \%}^{+9.8 \%}$ & $-0.001(20)_{-0 \%}^{+0 \%}~_{-0\%}^{+0\%}$ & $-0.10(3)_{-11.9 \%}^{+12.0 \%}~_{-2.8  \%}^{+2.8  \%}$ & $-0.006(40)_{-0 \%}^{+0\%}~_{-0 \%}^{+0 \%}$ & $-0.14(5)_{-28.1 \%}^{+25.5 \%}~_{-2.8 \%}^{+2.8 \%}$ \\
$B_n^+$ & $-0.04(2)_{-1.4  \%}^{+1.5  \%}~_{-4.3  \%}^{+4.3  \%}$ & $-0.03(2)_{-31.5  \%}^{+42.4  \%}~_{-1.6  \%}^{+1.6  \%}$ & $0.07(4)_{-47.6  \%}^{+73.0  \%}~_{-2.5  \%}^{+2.5  \%}$ & $-0.03(4)_{-44.1  \%}^{+54.9  \%}~_{-1.5  \%}^{+1.5  \%}$ & $0.12(6)_{-32.8  \%}^{+42.0  \%}~_{-2.5  \%}^{+2.5  \%}$ \\
$B_r^+$ & $0.003(20)_{-16.6  \%}^{+15.1  \%}~_{-2.0  \%}^{+2.0  \%}$ & $-0.04(2)_{-62.5  \%}^{+55.4  \%}~_{-4.2  \%}^{+4.2  \%}$ & $-0.11(4)_{-18.3  \%}^{+14.5  \%}~_{-2.0  \%}^{+2.0  \%}$ & $-0.06(4)_{-38.3  \%}^{+40.3  \%}~_{-3.2  \%}^{+3.2  \%}$ & $-0.15(6)_{-14.6  \%}^{+14.6  \%}~_{-2.1  \%}^{+2.1  \%}$ \\
$B_k^-$ & $0.01(2)_{-9.2  \%}^{+8.6  \%}~_{-16.5  \%}^{+16.5  \%}$ & $0.04(2)_{-31.5  \%}^{+47.9  \%}~_{-4.7  \%}^{+4.7  \%}$ & $-0.13(3)_{-11.3  \%}^{+9.4  \%}~_{-7.3  \%}^{+7.3  \%}$ & $0.05(4)_{-31.8  \%}^{+39.5  \%}~_{-4.7  \%}^{+4.7  \%}$ & $-0.18(5)_{-20.5  \%}^{+16.8  \%}~_{-8.1  \%}^{+8.1  \%}$ \\
$B_n^-$ & $-0.07(2)_{-0.7  \%}^{+1.0  \%}~_{-0.9  \%}^{+0.9  \%}$ & $-0.05(2)_{-18.6  \%}^{+23.3  \%}~_{-2.5  \%}^{+2.5  \%}$ & $-0.05(4)_{-49.1  \%}^{+35.9  \%}~_{-2.2  \%}^{+2.2  \%}$ & $-0.04(4)_{-35.1  \%}^{+36.8  \%}~_{-3.4  \%}^{+3.4  \%}$ & $-0.04(6)_{-100.2  \%}^{+88.0 \%}~_{-2.4  \%}^{+2.4  \%}$ \\
$B_r^-$ & $-0.01(2)_{-7.4  \%}^{+7.1  \%}~_{-0.6  \%}^{+0.6  \%}$ & $-0.02(2)_{-32.5  \%}^{+26.9  \%}~_{-3.0  \%}^{+3.0  \%}$ & $-0.13(4)_{-26.1  \%}^{+13.5  \%}~_{-2.6  \%}^{+2.6  \%}$ & $-0.02(4)_{-26.3  \%}^{+22.9  \%}~_{-2.0  \%}^{+2.0  \%}$ & $-0.17(6)_{-13.9  \%}^{+9.3  \%}~_{-3.7  \%}^{+3.7  \%}$ \\
\hline
\end{tabular}
\end{adjustbox}
\end{center}
\caption{The $B$ coefficients as defined in Eq.~\ref{eq:coeffdef} at various orders both expanded and unexpanded.}  
\label{table:coeff_B}
\end{table*}

\begin{table*}[htb!]
\begin{center}
\renewcommand{\arraystretch}{1.5}
\scriptsize
\begin{adjustbox}{width=1\textwidth}
\begin{tabular}{c c @{\hskip 20pt} r l @{\hskip 20pt} r l }
\hline
 & & \multicolumn{2}{c}{Unexpanded}  &  \multicolumn{2}{c}{Expanded}  \\
 &  LO QCD [\%] & NLO QCD [\%]  & NLO QCD+EW [\%]  &  NLO QCD [\%]  & NLO QCD+EW [\%] \\
\hline
$C_{kk}$ & $32.68(3)_{-1.7  \%}^{+1.5  \%}~_{-0.8  \%}^{+0.8  \%}$ & $32.88(3)_{-0.4  \%}^{+1.3  \%}~_{-0.7  \%}^{+0.7  \%}$ & $32.69(5)_{-0.4  \%}^{+1.1  \%}~_{-0.7  \%}^{+0.7  \%}$ & $32.96(7)_{-0.8  \%}^{+1.7  \%}~_{-0.6  \%}^{+0.6  \%}$ & $32.65(9)_{-0.6  \%}^{+1.5  \%}~_{-0.6  \%}^{+0.6  \%}$ \\
$C_{nn}$ & $33.01(3)_{-0.5  \%}^{+0.3  \%}~_{-0.2  \%}^{+0.2  \%}$ & $31.97(3)_{-1.1  \%}^{+0.9  \%}~_{-0.2  \%}^{+0.2  \%}$ & $31.89(5)_{-1.3  \%}^{+0.9  \%}~_{-0.2  \%}^{+0.2  \%}$ & $31.59(7)_{-0.5  \%}^{+0.6  \%}~_{-0.2  \%}^{+0.2  \%}$ & $31.49(9)_{-0.6  \%}^{+0.7  \%}~_{-0.2  \%}^{+0.2  \%}$ \\
$C_{rr}$ & $0.71(3)_{-51.5  \%}^{+45.1  \%}~_{-3.2  \%}^{+3.2  \%}$ & $4.80(3)_{-19.2  \%}^{+28.9  \%}~_{-2.8  \%}^{+2.8  \%}$ & $4.83(5)_{-19.4  \%}^{+29.4  \%}~_{-2.5  \%}^{+2.5  \%}$ & $6.28(7)_{-8.3  \%}^{+12.6  \%}~_{-1.9  \%}^{+1.9  \%}$ & $6.33(9)_{-8.1  \%}^{+11.6  \%}~_{-1.7  \%}^{+1.7  \%}$ \\
$C_{nr}+C_{rn}$ & $-0.02(4)_{-18.8  \%}^{+18.4  \%}~_{-1.0  \%}^{+1.0  \%}$ & $0.002(50)_{-0 \%}^{+0 \%}~_{-0  \%}^{+0  \%}$ & $0.08(7)_{-59.7  \%}^{+60.6  \%}~_{-3.8  \%}^{+3.8  \%}$ & $0.01(10)_{-0  \%}^{+0  \%}~_{-0  \%}^{+0  \%}$ & $0.1(1)_{-48.4  \%}^{+45.2  \%}~_{-3.6  \%}^{+3.6  \%}$ \\
$ C_{nr}-C_{rn} $ & $0.02(4)_{-12.0  \%}^{+10.5  \%}~_{-1.8  \%}^{+1.8  \%}$ & $-0.05(4)_{-53.6  \%}^{+29.9  \%}~_{-4.0  \%}^{+4.0  \%}$ & $-0.08(7)_{-54.3  \%}^{+34.1  \%}~_{-8.3  \%}^{+8.3  \%}$ & $-0.07(9)_{-31.6  \%}^{+22.1  \%}~_{-2.4  \%}^{+2.4  \%}$ & $-0.1(1)_{-28.3  \%}^{+23.0  \%}~_{-7.5  \%}^{+7.5  \%}$ \\
$C_{nk}+C_{kn}$ & $-0.02(4)_{-10.3  \%}^{+13.1  \%}~_{-3.3  \%}^{+3.3  \%}$ & $-0.004 (60)_{-0  \%}^{+0  \%}~_{-0  \%}^{+0  \%}$ & $-0.14(9)_{-40.6  \%}^{+30.8  \%}~_{-5.1  \%}^{+5.1  \%}$ & $0.001 (100)_{-0  \%}^{+0  \%}~_{-0  \%}^{+0  \%}$ & $-0.2(2)_{-30.3  \%}^{+28.6  \%}~_{-5.7  \%}^{+5.7  \%}$ \\
$C_{nk}-C_{kn}$ & $-0.004(40)_{-59.2  \%}^{+54.1  \%}~_{-3.3  \%}^{+3.3  \%}$ & $-0.03(6)_{-80.1  \%}^{+79.7  \%}~_{-13.9  \%}^{+13.9  \%}$ & $0.12(9)_{-11.4  \%}^{+13.7  \%}~_{-4.0  \%}^{+4.0  \%}$ & $-0.04(10)_{-83.5  \%}^{+78.4  \%}~_{-9.3  \%}^{+9.3  \%}$ & $0.2(2)_{-8.3  \%}^{+6.5  \%}~_{-4.1  \%}^{+4.1  \%}$ \\
$C_{rk}+C_{kr}$ & $-22.87(4)_{-1.5  \%}^{+1.5  \%}~_{-0.4  \%}^{+0.4  \%}$ & $-20.51(6)_{-2.8  \%}^{+3.8  \%}~_{-0.4  \%}^{+0.4  \%}$ & $-20.48(9)_{-2.9  \%}^{+4.1  \%}~_{-0.4  \%}^{+0.4  \%}$ & $-19.7(1)_{-1.4  \%}^{+2.3  \%}~_{-0.3  \%}^{+0.3  \%}$ & $-19.6(1)_{-1.5  \%}^{+2.2  \%}~_{-0.4  \%}^{+0.4  \%}$\\
$C_{rk}-C_{kr}$ & $0.02(4)_{-15.7  \%}^{+17.2  \%}~_{-5.1  \%}^{+5.1  \%}$ & $-0.13(6)_{-30.7  \%}^{+18.1  \%}~_{-2.2  \%}^{+2.2  \%}$ & $0.09(9)_{-29.4  \%}^{+52.4  \%}~_{-4.7  \%}^{+4.7  \%}$ & $-0.2(1)_{-13.1  \%}^{+10.2  \%}~_{-2.2  \%}^{+2.2  \%}$ & $0.1(2)_{-25.0  \%}^{+36.8  \%}~_{-4.8  \%}^{+4.8  \%}$ \\
\hline
\end{tabular}
\end{adjustbox}
\end{center}
\caption{The $C$ coefficients as defined in Eq.~\ref{eq:coeffdef} at various orders both expanded and unexpanded.}  
\label{table:coeff_C}
\end{table*}

\section{Results}
\label{sec:results}

We move now into presenting our results on the spin correlation coefficients, the asymmetries and the differential distributions.
\subsubsection*{Correlation coefficients}

We start with the Tabs. \ref{table:coeff_B}, \ref{table:coeff_C}, where we show the $B$ and $C$ spin correlation coefficients respectively. For the NLO QCD and the complete-NLO calculations we present the results from both the unexpanded and expanded definitions. In all cases we show the absolute statistical error and the relative scale and PDF uncertainties. \\

\noindent
In Tab.~\ref{table:coeff_B} we see that almost all of the coefficients are compatible with zero at LO and NLO QCD accuracy for both the unexpanded and expanded definitions. At the complete-NLO accuracy there is a small deviation from zero for some coefficients. In Tab.~\ref{table:coeff_C} for the $C$ coefficients, we find a similar trend for the coefficients being close to zero at LO, and remaining close to zero for most of the coefficients, except the $C_{kk}, C_{nn}, C_{rr}, C_{rk}+C_{kr}$ also at NLO QCD. For the vanishing $C$ coefficients, they remain compatible with zero also when the complete-NLO corrections are considered, in contrast to some of the $B$ coefficients which obtain a finite correction. For the four non-zero coefficients, the EW corrections are small and do not alter the NLO QCD prediction. They further show stability between the unexpanded and expanded predictions. \\

\noindent
In our calculation we do not include the NLO corrections to the top quark decays and the spin correlations from the virtual part of the NLO in the $t \bar t$ production. These effects can be evaluated by comparison of the results in Ref.~\cite{Czakon:2020qbd}, which include NLO QCD corrections in both the production and decay, to the results obtained via MCFM based on Ref.~\cite{Campbell:2012uf}, where the decay is included at LO. We list in Tab. \ref{tab:comparison} the values from Ref.~\cite{Czakon:2020qbd}, and the results one obtains from the MCFM implementation for the four non-zero spin correlation coefficients.
\begin{table}[h]
\begin{center}
\renewcommand{\arraystretch}{1.5}
\scriptsize
\begin{tabular}{c @{\hskip 20pt} c c c c }
\hline
QCD order / [\%] &  $C_{kk}$ &  $C_{nn}$ & $C_{rr}$  &  $C_{rk+kr}$ \\
\hline
NLO $\times$ NLO (\cite{Czakon:2020qbd} ) &  33.0(3) &33.0(2) & 5.8(2) & -20.3(2) \\
NLO $\times$ LO (MCFM) &  33.04(4) & 33.09(4) & 5.96(4) & -20.71(7) \\
\hline
\end{tabular}
\end{center}
\caption{The four largest spin correlation coefficients at NLO in decay (from Ref.~\cite{Czakon:2020qbd}) and with LO in decay based on the MCFM parton-level process library.}  
\label{tab:comparison}
\end{table}
\noindent

\begin{table*}[!htbp]
\begin{center}
\renewcommand{\arraystretch}{1.5}
\scriptsize
\begin{adjustbox}{width=1\textwidth}
\begin{tabular}{c c @{\hskip 20pt} r l @{\hskip 20pt} r l }
\hline
 & & \multicolumn{2}{c}{Unexpanded}  &  \multicolumn{2}{c}{Expanded}  \\
 Asymmetry &  LO QCD [\%] & NLO QCD [\%] & NLO QCD+EW [\%] & NLO QCD [\%] & NLO QCD+EW [\%] \\
 \hline
$A_{C}^{tt} $ & 0 & $0.453(5)_{-20.5  \%}^{+28.2  \%}~_{-3.4  \%}^{+3.4  \%}$ & $0.546(6)_{-18.0  \%}^{+25.1  \%}~_{-2.4  \%}^{+2.4  \%}$ & $0.62(2)_{-14.8  \%}^{+18.1  \%}~_{-3.3  \%}^{+3.3  \%}$ & $0.73(3)_{-11.5  \%}^{+13.8  \%}~_{-2.3  \%}^{+2.3  \%}$ \\
$A_C^{\ell \ell} $ & 0 & $0.27(2)_{-21.4  \%}^{+29.3  \%}~_{-3.8  \%}^{+3.8  \%}$ & $0.33(3)_{-17.8  \%}^{+25.0  \%}~_{-3.8  \%}^{+3.8  \%}$ & $0.36(3)_{-15.9  \%}^{+19.3  \%}~_{-3.7  \%}^{+3.7  \%}$ & $0.45(4)_{-12.0  \%}^{+14.6  \%}~_{-3.9  \%}^{+3.9  \%}$ \\
$A_{\Delta\Phi}$ & $17.51(1)_{-2.8  \%}^{+3.2  \%}~_{-0.4  \%}^{+0.4  \%}$ & $12.65(2)_{-14.8  \%}^{+8.3  \%}~_{-0.4  \%}^{+0.4  \%}$ & $12.42(3)_{-15.5  \%}^{+8.7  \%}~_{-0.4  \%}^{+0.4  \%}$ & $10.88(3)_{-10.1  \%}^{+7.2  \%}~_{-0.3  \%}^{+0.3  \%}$ & $10.58(4)_{-10.5  \%}^{+7.4  \%}~_{-0.4  \%}^{+0.4  \%}$ \\
$A_{\Delta\theta}$ & $14.63(1)_{-4.6  \%}^{+4.0  \%}~_{-1.5  \%}^{+1.5  \%}$ & $16.03(2)_{-2.2  \%}^{+4.0  \%}~_{-1.4  \%}^{+1.4  \%}$ & $16.24(2)_{-2.2  \%}^{+4.1  \%}~_{-1.4  \%}^{+1.4  \%}$ & $16.54(3)_{-1.7  \%}^{+2.9  \%}~_{-1.3  \%}^{+1.3  \%}$ & $16.83(4)_{-1.5  \%}^{+2.8  \%}~_{-1.3  \%}^{+1.3  \%}$ \\
\hline
\end{tabular}
\end{adjustbox}
\end{center}
\caption{Asymmetries at various orders as defined in Sec. \ref{sec:asymmetry}, both expanded and unexpanded.}  
\label{table:asymm}
\end{table*}

As can be seen from this table, the NLO QCD corrections in the decay do not affect these spin correlation coefficients. This allows a safe conclusion that our results for the spin correlations with complete NLO in production and LO in decay should not be greatly altered if one includes also the NLO QCD in the decay. The differences between the coefficients obtained with MCFM in Tab. \ref{tab:comparison} and our corresponding results in Tab. \ref{table:coeff_C} is of the order of percent level with the exception of the $C_{rr}$, where they are at $\sim\!\!25\%$. The origin of these differences are the virtual spin corrections, which are approximated with the tree-level ones in our calculation, but are included in the MCFM calculation. Both the results obtained via MCFM and the results in Ref.~\cite{Czakon:2020qbd} agree with our results regarding the spin correlation coefficients that are compatible with zero.

\subsubsection*{Asymmetries}

Moving to the asymmetries shown in Tab.~\ref{table:asymm}, in the first line we show the already known $A_{C}^{tt}$, which is in agreement with the results obtained from the $t\bar t$ production asymmetry study \cite{Czakon:2017lgo} for the corresponding orders at 13 TeV. We further present the dilepton asymmetries as defined in Sec. \ref{sec:asymmetry}. The $A_C^{\ell \ell}$ is smaller in comparison to the $A_{C}^{tt}$ at all perturbative orders, and the complete NLO has a similar effect, an increase of $\sim 20\%$ for both of the asymmetries as compared to the NLO QCD. The $A_{\Delta\Phi}$, $A_{\Delta\theta}$ are non-zero already at LO QCD. For the $A_{\Delta\Phi}$ the NLO QCD corrections reduce the asymmetry and the EW ones further reduce it but by a smaller amount. The $A_{\Delta\theta}$ behaves in the opposite way regarding the higher order corrections. In both cases the expanded definitions behave accordingly.

\subsubsection*{Differential distributions}

Regarding the top quark pair distributions ($p_T(t), m(t \bar t), y(t)$) we have checked that in our results the effects of the EW corrections are in agreement with the results presented in Ref.~\cite{Czakon:2017wor}.\\

\begin{figure*}[ht!]
\centering
\begin{multicols}{2}
\includegraphics[width=0.9\linewidth]{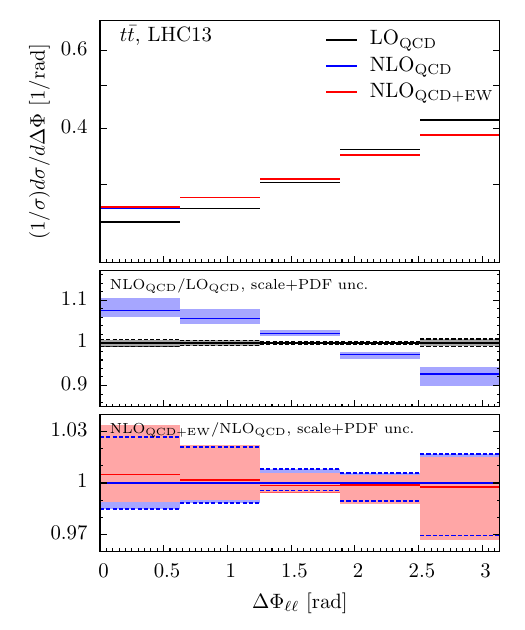} \par
\includegraphics[width=0.9\linewidth]{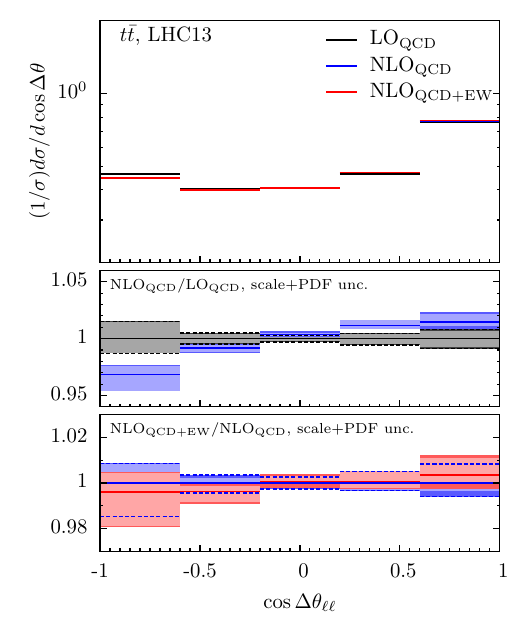} \par
\end{multicols}
\begin{multicols}{2}
\includegraphics[width=0.9\linewidth]{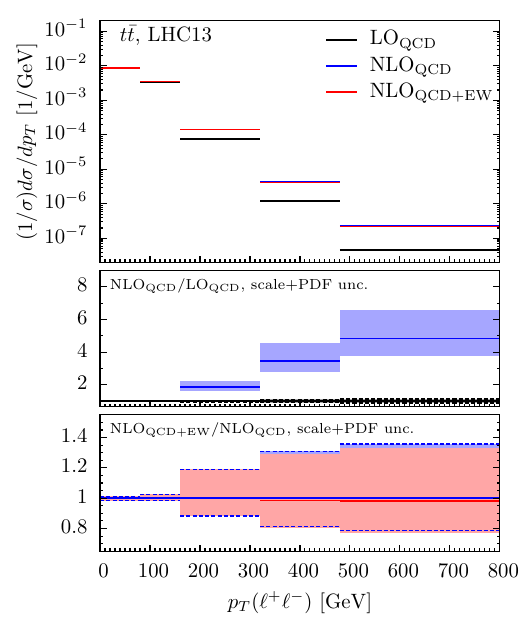}\par
\includegraphics[width=0.9\linewidth]{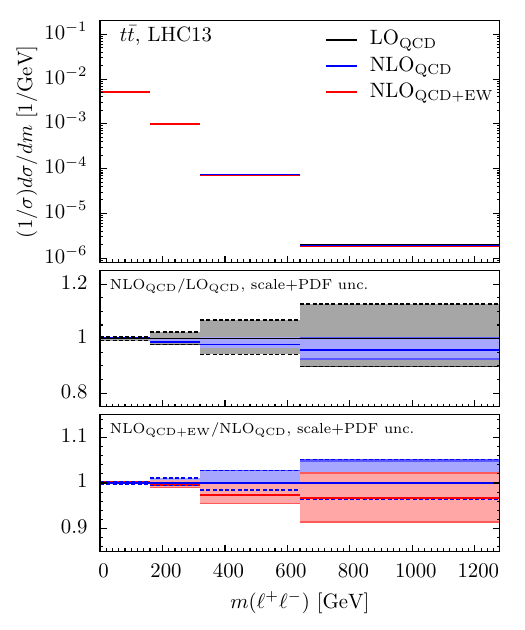}\par
\end{multicols}
\caption{The $\Delta\Phi_{\ell\ell}$ distribution (upper left) and $\cos\Delta\theta_{\ell\ell}$ distributions (upper right), transverse momentum of $\ell^+\ell^-$ system (lower left) and dilepton invariant mass distribution (lower right).}
\label{fig:lepton_pair}
\end{figure*}

\begin{figure*}[h!]
\centering
\begin{multicols}{2}
\includegraphics[width=0.9\linewidth]{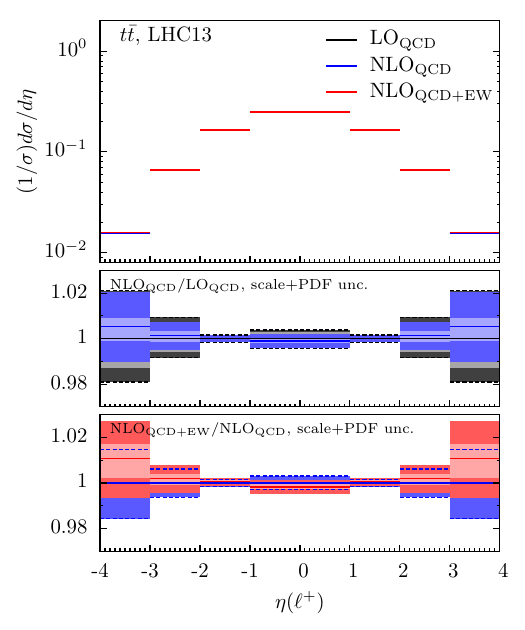}\par
\includegraphics[width=0.9\linewidth]{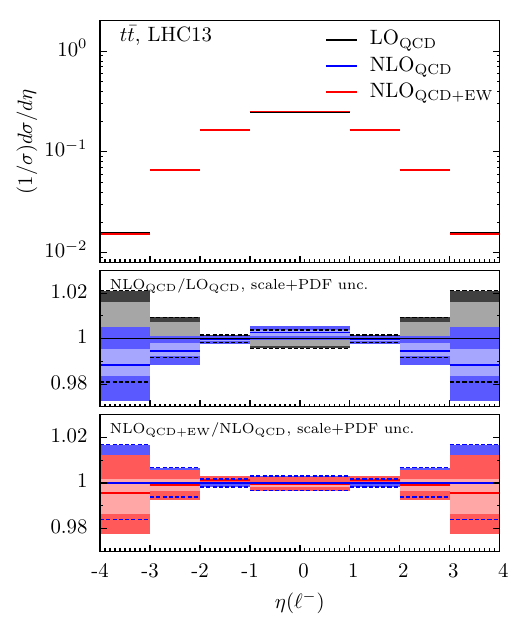}\par
\end{multicols}
\begin{multicols}{2}
\includegraphics[width=0.9\linewidth]{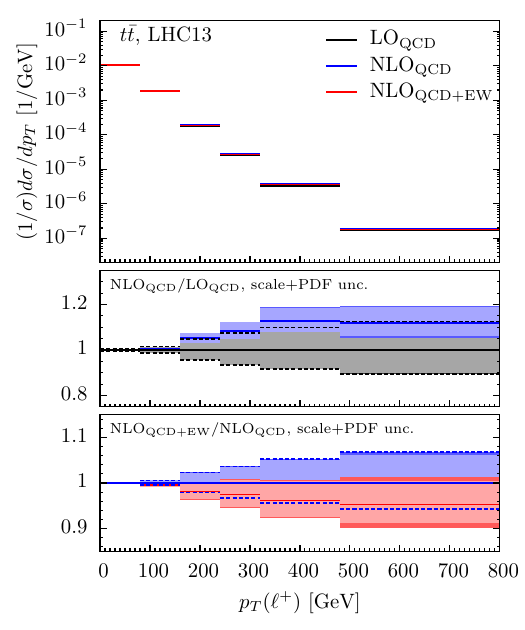}\par
\includegraphics[width=0.9\linewidth]{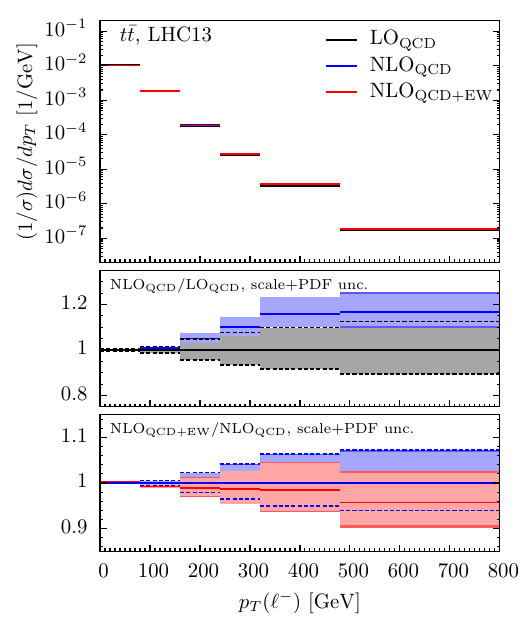} \par
\end{multicols}
\caption{Pseudorapidity (upper) and transverse momentum (lower) lepton distributions.}
\label{fig:sep_leptons}
\end{figure*}

\noindent
For the decay level differential distributions we focus on the leptons. We show the normalised leptonic distributions in Figs.~\ref{fig:lepton_pair}, \ref{fig:sep_leptons}, since they significantly reduce the theoretical uncertainties as well as the systematic experimental errors with respect to the unnormalised ones. In all cases we show the LO QCD, NLO QCD and NLO QCD+EW distributions. In the first inset we show the ratio between the NLO QCD and LO QCD normalised differential distributions. In the second inset we present the NLO QCD+EW over the NLO QCD one. \\

\noindent
In both insets we show the scale and PDF uncertainties for all the predictions. In all cases the scale uncertainties are the light-coloured band around the central value and the added PDF uncertainties in quadrature to the scale ones are the dark-coloured band. We start our discussion with the lepton pair angular distributions $\Delta\Phi_{\ell\ell}$ and $\cos\Delta\theta_{\ell\ell}$ in Fig.~\ref{fig:lepton_pair} (upper plots), which are directly connected to the corresponding asymmetries presented in Tab. \ref{table:asymm}.
In the first inset of the $\Delta\Phi_{\ell\ell}$ and $\cos\Delta\theta_{\ell\ell}$ plots we see the large effect of the QCD corrections. Furthermore the shape of the QCD corrections is in agreement with the reduction and increase of the $A_{\Delta\Phi}$ and $A_{\Delta\theta}$ respectively, as shown in Tab. \ref{table:asymm}, since the dilepton asymmetries $A_C^{\ell \ell}, A_{\Delta\Phi}, A_{\Delta\theta}$ are fully correlated to the leptonic differential distributions $\eta(\ell), \Delta\Phi_{\ell\ell}$ and $\cos\Delta\theta_{\ell\ell}$. In the second inset of these plots we see the small effect (below one percent) of the EW corrections in both distributions.\\

\noindent
We continue the description of the lepton pair by showing the invariant mass and the transverse mass of the system in Fig. \ref{fig:lepton_pair} (lower plots). In the $p_T(\ell^+\ell^-)$ we firstly observe the very large effect of the NLO QCD corrections, a giant $K$-factor appearing in the highly cross-section suppressed tail, which is already known~\cite{Czakon:2020qbd}, and secondly that the effect of the EW corrections is negligible for this distribution. Moving to the $m(\ell^+\ell^-)$ distribution we see an EW effect of $\sim2\%$ below 500 GeV and of $\sim5\%$ after 500 GeV. The $m(\ell^+\ell^-)$ spectrum is softened with respect to the NLO QCD prediction. The effect lies along the lower band of the NLO QCD normalised uncertainties, but given the fact that the QCD predictions have reached the NNLO accuracy \cite{Czakon:2020qbd} with the scale uncertainty being decreased to a few percent level, the EW corrections will alter the prediction for this observable. For the distributions shown in Fig. \ref{fig:lepton_pair} the PDF uncertainties are very small to negligible w.r.t. the scale uncertainties.\\

\noindent
Regarding the separate lepton distributions we show the pseudorapidities and the transverse momenta of the leptons emerging from top quark pair in Fig. \ref{fig:sep_leptons}. In the pseudorapidity distributions (upper plots) we see in the first insets the non-zero NLO QCD induced lepton asymmetry, which is also shown in Tab. \ref{table:asymm}. In the second insets we see the enhancement of this feature due to the EW corrections, but in both cases these effects are below $\sim 2\%$ and appear at large rapidity regimes ($|\eta|>3$). From the shape of the ratio insets we can see that the observed effects from both the QCD and EW corrections will cancel in the case of an averaged $\eta(\ell)$ distribution. In the transverse momentum distributions (lower plots) we see the large positive effect of the NLO QCD corrections in the first inset. In the second inset we observe that the EW effects become negative at higher $p_T$ values reaching a negative correction of $\sim 5\%$ with respect to the NLO QCD prediction between 500 and 800 GeV. In the case of the lepton $p_T$'s as well as the $m(\ell^+ \ell^-)$ the EW corrections bring the predictions closer to the data at the tails, according to the comparisons between NNLO QCD predictions and experimental results in Ref.~\cite{Czakon:2020qbd}. The high $|\eta(\ell)|$ regimes show scale and PDF uncertainties of the same level, whereas the theory uncertainties of the $p_T(\ell)$ distributions are fully dominated by the scale variation. Our results concerning the $\Delta\Phi_{\ell\ell},\cos\Delta\theta_{\ell\ell}$ and $p_T(\ell)$ distributions are qualitatively similar to what has been found in Ref.~\cite{Denner:2016jyo}, where part of the NLO EW corrections is included.

\section{Conclusion and discussion}
\label{sec:conclusion}

In this work we have presented for the first time the complete-NLO corrections of the $t \bar t$ production to the spin correlation coefficients, the leptonic asymmetries and the differential leptonic distributions. We have focused on the process of top quark pair production and decay to a dilepton channel at fixed-order. We tested our method for the reweighting at fixed-order by comparing the $t\overline{t}$ asymmetry and top quark distributions to already known results and found that they are in good agreement.\\

\noindent
Our calculation has two major approximations: firstly, we consider the tree-level spin correlations. This was investigated by comparison to available results in the literature to find that one of the four dominant spin-correlation coefficients obtain a sizeable difference, while the other coefficients are insensitive to the virtual spin correlations. Secondly, we use an accuracy of electroweak precision in the production, while keeping a LO precision in the decay. Despite these approximations, the present results indicate the size of the electroweak effects. \\

\noindent
We investigated in particular the effects on the spin correlation in terms of the expansion coefficients within the spin-density formalism. We found that the EW effects, now included in the complete NLO, have very small effect compared to the NLO QCD results on most of the spin correlation coefficients. However, for some coefficients the contribution moves the result from zero to a finite value. We investigated the coefficients both expanded and unexpanded in the strong coupling, and found similar effects of the complete NLO in both versions. We further investigated the asymmetries, both those which have been known from previous calculations (the $t\overline{t}$ central-peripheral asymmetry) and the leptonic asymmetries calculated for the first time at this accuracy. We found there that the EW effects are present with a few percent effect. Finally we have studied the normalised leptonic differential distributions. We found that in specific phase-space regions of the transverse momentum and invariant mass distributions the complete-NLO calculation softens the NLO QCD prediction by a few percent, lying close to the lower border of the NLO QCD uncertainty band. 

\section*{Acknowledgements}

We thank Olivier Mattelaer for discussions regarding MadSpin. IT would like to thank Eleni Vryonidou for useful discussions on the spin correlation coefficient definitions. This work is done in the context of and supported by the Swedish Research Council under contract number 2016-05996. The work of IT is also supported by the MorePheno ERC grant agreement under number 668679. Computational resources to IT have been provided by the Consortium des \'Equipements de Calcul Intensif (C\'ECI), funded by the Fonds de la Recherche Scientifique de Belgique (F.R.S.-FNRS) under Grant No. 2.5020.11 and by the Walloon Region.

\bibliographystyle{hieeetr}
\bibliography{Article_draft} 

\end{document}